\documentclass[aps,prl,twocolumn,superscriptaddress]{revtex4-2}

\usepackage{graphicx}
\usepackage{dcolumn}
\usepackage{bm}
\usepackage{amsmath}
\usepackage{color}

\begin{document}


\title{{Fast Magnetosonic Turbulence in Two-Dimensional Relativistic Plasmas}
} 
\author{Petr Ugarov}
\author{Vladimir Zhdankin}
\author{Giuseppe Arrò}
\affiliation{
 Department of Physics, University of Wisconsin-Madison, Madison, WI 53706, USA
}
\date{\today}

\begin{abstract}
We present fully kinetic simulations of driven 2D turbulence in a relativistic plasma, designed for the first time to induce a fast magnetosonic cascade. As the driving strength increases, turbulence transitions from a weak wave-dominated regime to strong shock-driven dynamics. Using spatiotemporal Fourier analysis, we identify fast modes, finding that the weak turbulence regime exhibits spectral properties that are in excellent agreement with theoretical expectations. Our results are relevant for the modeling of turbulence in high-energy astrophysical plasmas.
\end{abstract}

\maketitle

{\it Introduction---} Compressible turbulence has proved to be a rich phenomenon in space and astrophysical plasmas, being relevant in a variety of contexts, from the Earth's magnetosheath (where fast magnetosonic modes play a crucial role \cite[][]{Zhao2024}) to distant supernova remnants \cite{Reynolds2008}. In some systems, such as pulsar wind nebulae and jets from active galactic nuclei, the plasma is relativistic \cite{Kirk2009, Ferrari2003}, with internal energy and magnetic fields dominating over the rest mass energy. Despite its importance, compressible turbulence is generally less understood than its incompressible counterpart. While some aspects of compressible turbulence are understood in the magnetohydrodynamic (MHD) framework \citep{Schekochihin2022}, the situation is less established for collisionless plasmas, where kinetic models are required. Fully kinetic particle-in-cell (PIC) simulations have been employed to investigate nonlinear wave interactions and shock formation in collisionless plasmas \cite{Svidzinski2009}, but a fast-mode–dominated turbulent cascade has not yet been isolated.

Consequently, the role of fast modes in mediating energy transfer within compressible turbulence remains unclear. MHD models predict that fast modes cascade to form approximately isotropic energy spectra \cite{ChoLazarian2002, Galtier2023}, similar to acoustic turbulence. Acoustic turbulence itself admits qualitatively distinct regimes of weak and strong turbulence. In the weak regime, wave interactions are perturbative and lead to the $k^{-3/2}$ Zakharov-Sagdeev spectra \cite{ZakharovSagdeev1970}, whereas in the strong regime nonlinear steepening leads to shock formation with a $k^{-2}$ Kadomtsev-Petviashvili spectrum \cite{KadomtsevPetviashvili1973, Burgers1948}. Recent theoretical and experimental studies have demonstrated the coexistence of weak and strong acoustic turbulence, along with transitions between them \cite{Kochurin2024, Chen2024}. Whether an analogous distinction applies to fast-mode turbulence in collisionless plasmas remains unclear. 

However, current MHD predictions of fast-mode turbulence neglect collisionless and relativistic effects that may qualitatively alter the fast-mode dynamics in high-energy plasmas. Recent kinetic simulations have begun to explore collisionless effects \citep{Hou2025} suggesting spectral indices close to $k^{-3/2}$ in the absence of strong damping. In 3D, simulating compressible kinetic turbulence at high Mach number remains prohibitively expensive on modern supercomputers. In contrast, lower dimensional studies offer significant advantages in computational and analytical simplicity.
Compressible supersonic turbulence has recently been simulated in 2D using a hybrid-kinetic model \citep{gootkin_etal_2025}, demonstrating nonthermal particle acceleration by the intermittent shocks. In addition, compressible supersonic turbulence has also been studied in a 3D hybrid-kinetic simulation of a turbulent dynamo \citep{chirakkara_etal_2025}. 
However, the case of a kinetic fast-mode-dominated cascade is computationally unexplored even in lower dimensional studies. Furthermore, all of these studies have so far neglected relativistic effects. Relativistic extensions of weak MHD turbulence have been studied \citep{Tenbarge2021, Gao2026}, but relativistic fast-mode kinetic turbulence remains unexplored.

In this Letter, we present the first fully kinetic PIC simulations of driven, 2D relativistic turbulence in a pair plasma designed to isolate a fast-mode–dominated cascade in the subsonic regime. Using spatiotemporal Fourier analysis, we directly identify fast-mode branches in the turbulent spectrum. We further demonstrate that the driving amplitude determines whether fast modes steepen into shocks or not, and we numerically determine spectral indices for spectra of fast-mode wave turbulence. 

{\it Methods---} We consider an electron-positron (pair) plasma that is ultra-relativistically hot, with temperatures $\theta \equiv T/ m_e c^2 \gg 1$ (normalized to the electron rest mass energy, where $m_e$ is the electron mass). We consider a 2D domain (in the $x$-$y$ plane) with an initially homogeneous magnetic field $\boldsymbol{B}_0 = B_0 \hat{\boldsymbol{z}}$, constant total particle number density $n_0$, and temperature $\theta_0$. No magnetic field, electric field and velocity fluctuations are present initially. When this setup is driven by in-plane forces (as done in our simulations), the in-plane magnetic field and the out-of-plane electric field always stay zero, and the out-of-plane particle momentum is conserved (i.e. the momentum-space is effectively 2D). 
Out of the large scale ideal MHD waves, only fast magnetosonic modes (FM) develop in this setup. Due to their wavevector $\boldsymbol{k}$ being perpendicular to $\boldsymbol{B}$, FMs experience very little collisionless damping unless $k_\perp \rho_e \gtrsim 1$ \cite{barnes_1966}, where $\rho_e = \langle p_\perp\rangle c/eB$ (where $e$ is the elementary charge) is the electron Larmor radius given the average in-plane momentum $\langle p_\perp \rangle$, and $k_\perp = (k_x^2+k_y^2)^{1/2}$ (with $k_x$ and $k_y$ being wavenumbers in the $x$ and $y$ direction). 

We perform a linear calculation of the plasma electric susceptibility in two different limits \citep{Stix1992}: the low-frequency, large-scale MHD limit, where $\omega/\Omega_e \ll 1$ and $k_\perp\rho_e \ll 1$, with $\omega$ being the wave frequency while $\Omega_e = eB/m_e \langle \gamma \rangle c$ is the electron gyrofrequency; and the high-frequency small-scale kinetic limit, where $\omega/\Omega_e \gg 1$ and $k_\perp\rho_e \gg 1$.
Assuming an ultrarelativistic Maxwell-Jüttner distribution as relevant for our simulations, the dispersion relations in the two limits are: 
\begin{eqnarray}
\omega(k_\perp) &=& \begin{cases} v_Fk_\perp \left(1 - A \frac{c^2}{\Omega_e^2} k_\perp^2\right) \, , \: \: \: {k_\perp\rho_e \ll 1} \\
ck_\perp \left(1 + \frac{3\omega_{pe}^2}{32 c^2 k_\perp^2}\right) \, , \: \: \: k_\perp\rho_e \gg 1
\end{cases} \, .
\label{eq:dispersion}
\end{eqnarray} 
Here
\begin{eqnarray}
A \equiv \frac{7/10 + (7/5) v_A^2/c^2 + (5/2) v_A^4/c^4}{3(1+v_A^2/c^2)^2 (v_A^2/c^2 + 2/5)} \, , \nonumber
\end{eqnarray}
$\omega_{pe} = \sqrt{4 \pi n_0 e^2/6m_e \theta}$  is the ultrarelativistic electron plasma frequency, $v_F \equiv [(v_A^2 + c_s^2)/(v_A^2/c^2 + 1)]^{1/2}$ is the relativistic FM velocity, $v_A = B/{\sqrt{8\pi n_0 \theta m_e}}$ is the Alfv\'{e}n velocity in the $\theta \gg 1$ limit, and $c_s = (2/5)^{1/2} c$ is the sound speed. Thus, the first limit in Eq.~\ref{eq:dispersion} corresponds to a relativistic FM, while the second limit is an electromagnetic wave. The FM velocity differs from the MHD prediction, due to kinetic effects modifying the sound speed. 

We perform a series of numerical simulations of externally driven turbulence, solving the Vlasov-Maxwell equations with the PIC code {\em Zeltron} \cite{Cerutti2013}. The simulation domain is a 2D periodic box of area $L^2$. The electrons and positrons are initialized with the uniform Maxwell-Jüttner distribution having an initial temperature $\theta_0 = 10$.  

We use an external driver $\boldsymbol{F}_{\rm ext}$ to produce large-scale velocity fluctuations, as described in Ref.~\cite{zhdankin_2021}. $\boldsymbol{F}_{\rm ext}$ is purely compressive, i.e. $\nabla_\perp \times \boldsymbol{F}_{\rm ext} = 0$, and is formed by a superposition of randomly evolved Fourier modes with wavenumbers $k \le 6\pi/L$, frequencies $\omega_k = 0.5 v_F k$, and decorrelation rates $\Gamma_k = 0.8 \omega_k$. We vary the non-dimensionalized driving force $F \equiv 8 N_k^{1/2} (n_0 L/B_0^2) |\boldsymbol{F}^k_{\rm ext}|$, where $N_k = 14$ is the number of driven wavevectors and $\boldsymbol{F}^k_{\rm ext}$ is the characteristic amplitude per mode. 

We perform a parameter scan of the non-dimensionalized driving force $F \in \{ 1/8, 1/4, 1/2, 1, 2, 4 \}$. The domain size is $L/\rho_e = 512$, sampled by a uniform periodic grid with $1536^2$ cells, and $32$ particles-per-cell for each species. We also perform a simulation with $F=1/4$, having 160 particles-per-cell and $L/\rho_e = 256$ ($768^2$ cells), that we use to analyze spatiotemporal spectra. We set the initial plasma beta $\beta_0 = 8\pi n_0 \theta m_ec^2/B_0^2 =1$ for all simulations described in this Letter. For these parameters, we get $v_F/c$ = 0.77. In addition, we performed simulations with a lower $\beta_0 = 0.25$, which yielded qualitatively similar results in the inertial range, as expected since $v_F$ has a weak dependence on $\beta$.

{\it Results---} 
Compressive driving is effective for generating FMs, which can steepen into shocks in supersonic turbulence \cite{ChoLazarian2003}. We characterize the presence of shocks in our simulations using a parameter called the supersonic fraction $S$, defined as the fraction of the domain that is supersonic, i.e. having local fast magnetosonic Mach numbers $M = |\boldsymbol{u}|/v_F \geq 1$, where $\boldsymbol{u}$ is the local fluid flow velocity defined in the Eckart frame. Shocks are absent if $S \ll 1$.

\begin{figure}[t]
\includegraphics[width=\linewidth]{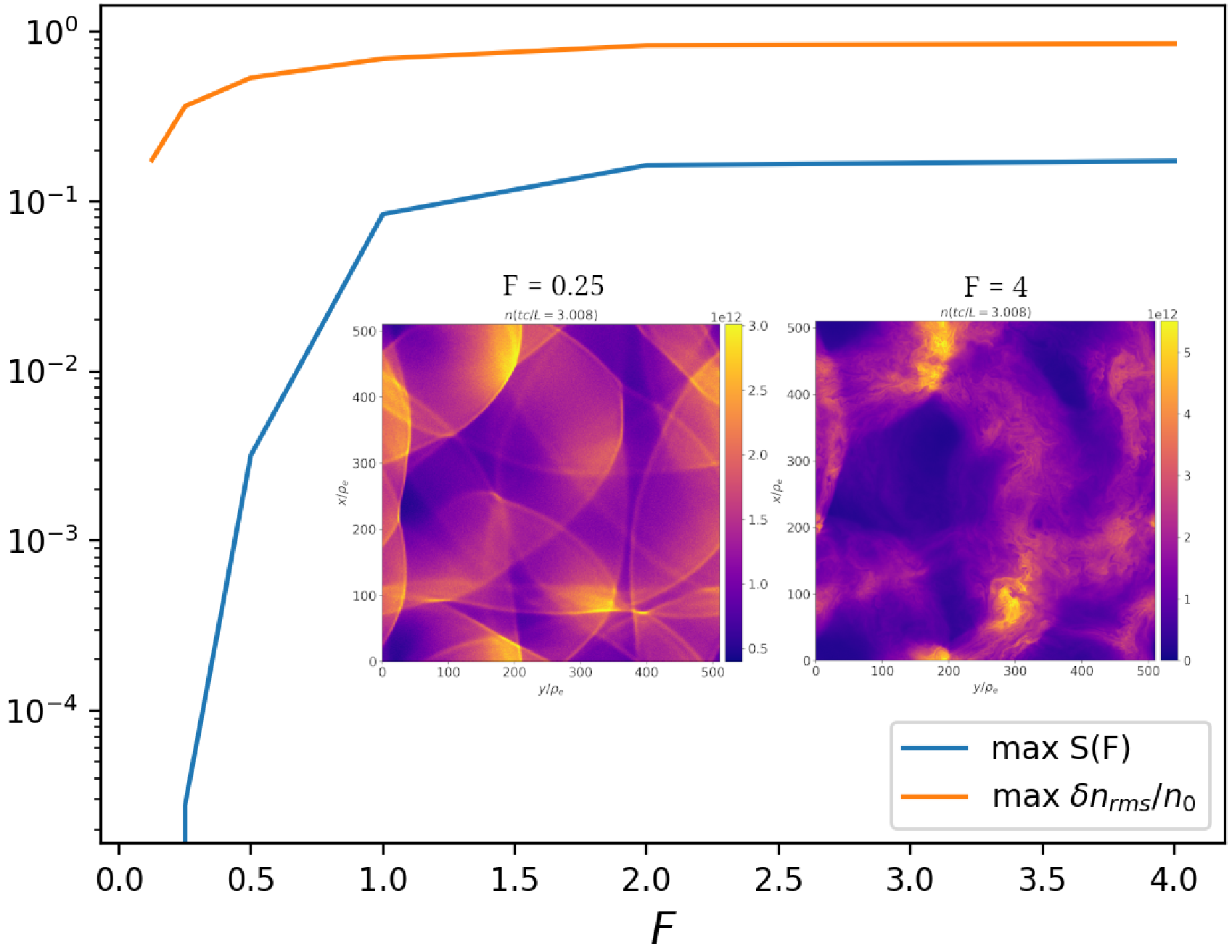}
\caption{The temporal maximum of the supersonic fraction, $\text{max}(S)$ (blue), and rms density fluctuations, $\text{max}(\delta n_{\rm rms}/n_0)$ (orange), versus the driving strength $F$. The insets show a density snapshot of the simulations at a time of approximately $3L/c$ at different driving strengths $F$ of 0.25 and 4, respectively.}
\label{fig:shockness}
\end{figure}

For each simulation, we measure $S$ as a function of time; we find that $S$ starts near zero, reaches a peak after approximately a light crossing time $t \approx L/c$, and then slowly declines in time. In Fig.~\ref{fig:shockness}, we show the temporal maximum of the supersonic fraction, $\max{S}$, for the simulations with varying $F$ (blue curve). 
We find that $\max{S}$ strongly increases with driving strength $F$, until it saturates at $F \approx 1$. Density snapshots of both $F = 0.25$ and $F = 4$ simulations at $tc/L \approx 3$ are also shown in Fig.~\ref{fig:shockness}, showing that the weakly driven regime consists of smooth and coherent density waves, while the strongly driven regime has a much more irregular morphology. The density animation movies for these two cases can be seen in Supplemental Material \cite{SuppMaterial}. In addition, we show the maximum rms density fluctuations $\text{max}(\delta n_{\rm rms}/n_0)$ (orange curve) in each simulation, with $\delta n_{\rm rms} = \sqrt{\langle n^2\rangle}$. We can see that the density fluctuations increase with driving strength, as expected, but eventually saturate at $\text{max}(\delta n_{\rm rms}/n_0) \sim 1.$  

\begin{figure*}
  \centering
  \includegraphics[width=0.49\textwidth]{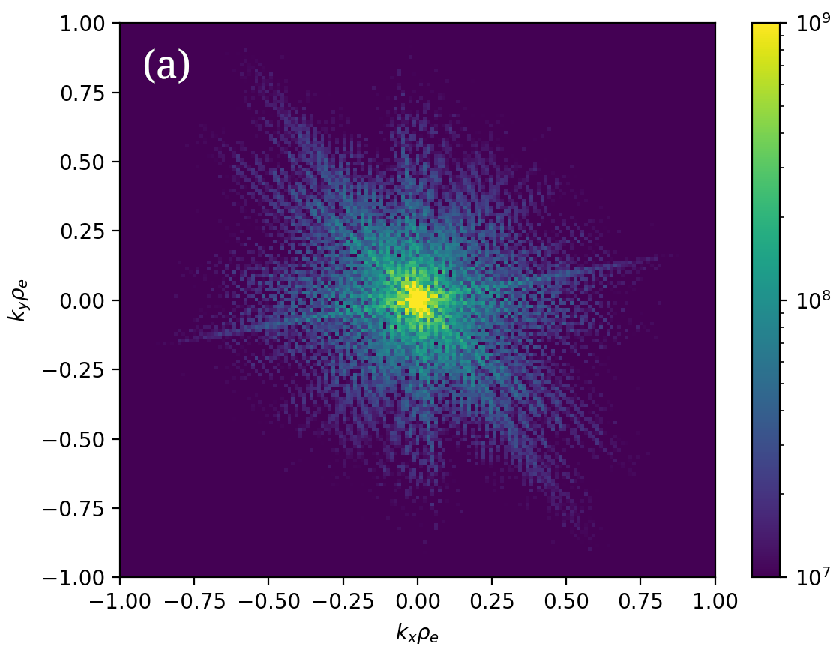}
  \hfill
  \includegraphics[width=0.49\textwidth]{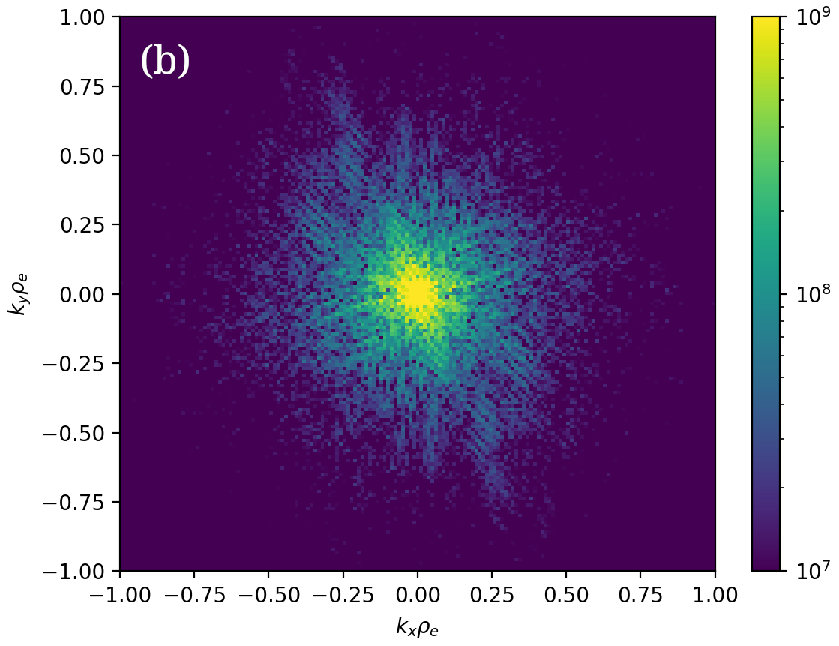}
  \caption{Two-dimensional Fourier power spectra of $B(x,y)$ in the $k_x-k_y$ plane at $t  \approx 3.8L/c.$ Panel (a) shows the weak-driving case ($F = 0.25$), while (b) corresponds to the strong-driving regime ($F = 2$).}
  \label{fig:hedgehog}
\end{figure*}

\begin{figure*}
  \centering
  \includegraphics[width=0.32\textwidth]{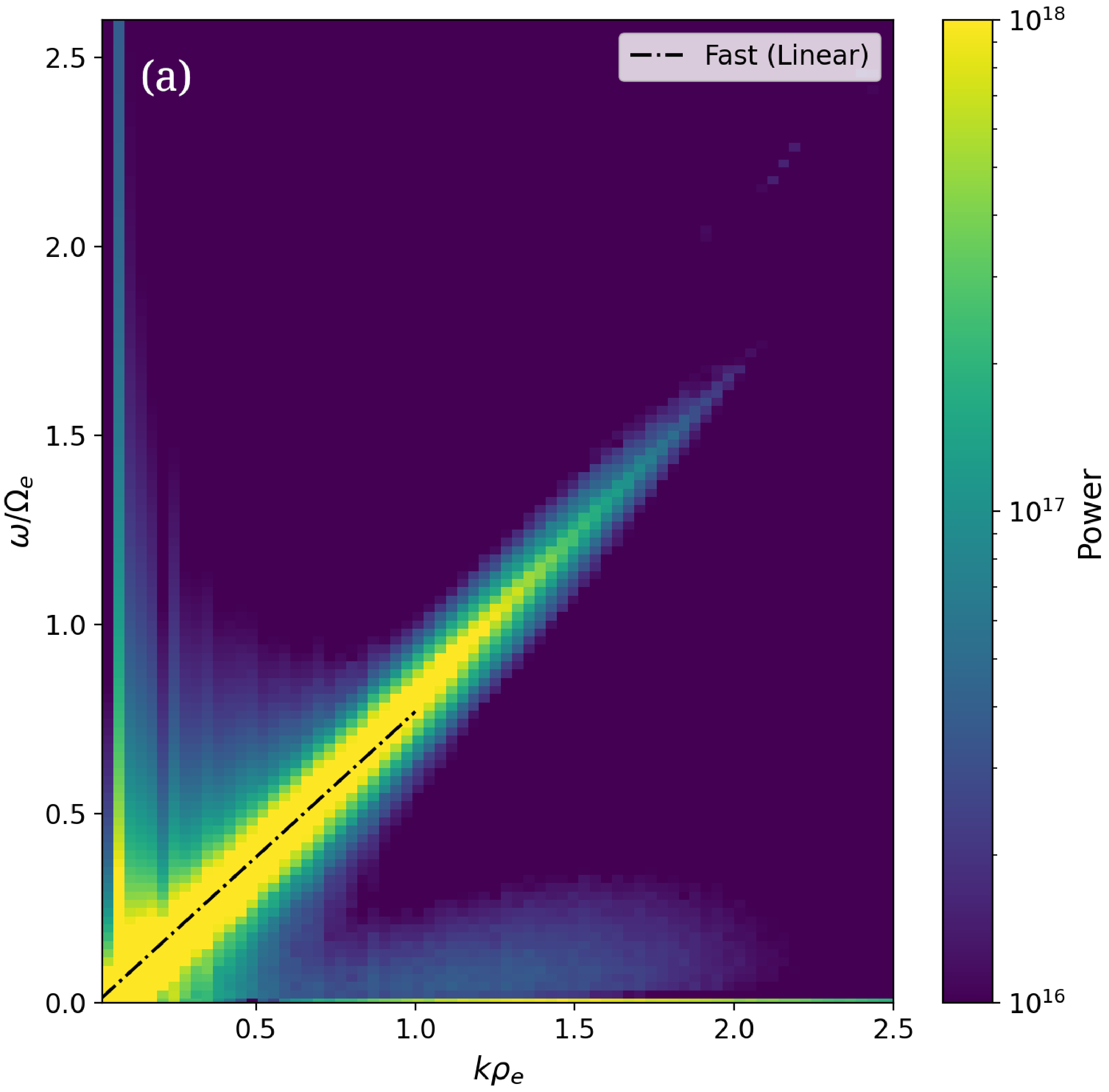}
  \hfill
  \includegraphics[width=0.32\textwidth]{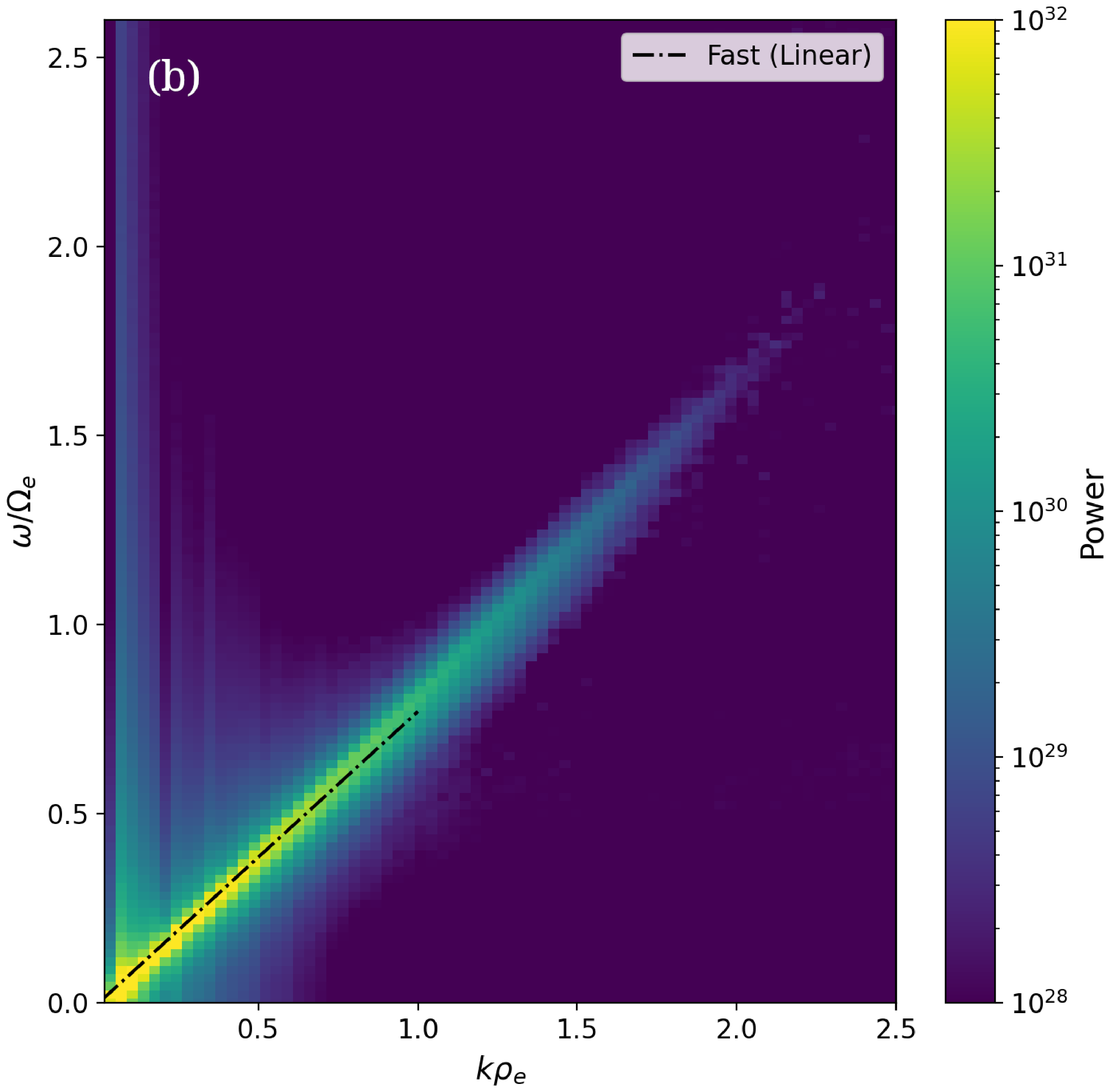}
  \hfill
  \includegraphics[width=0.32\textwidth]{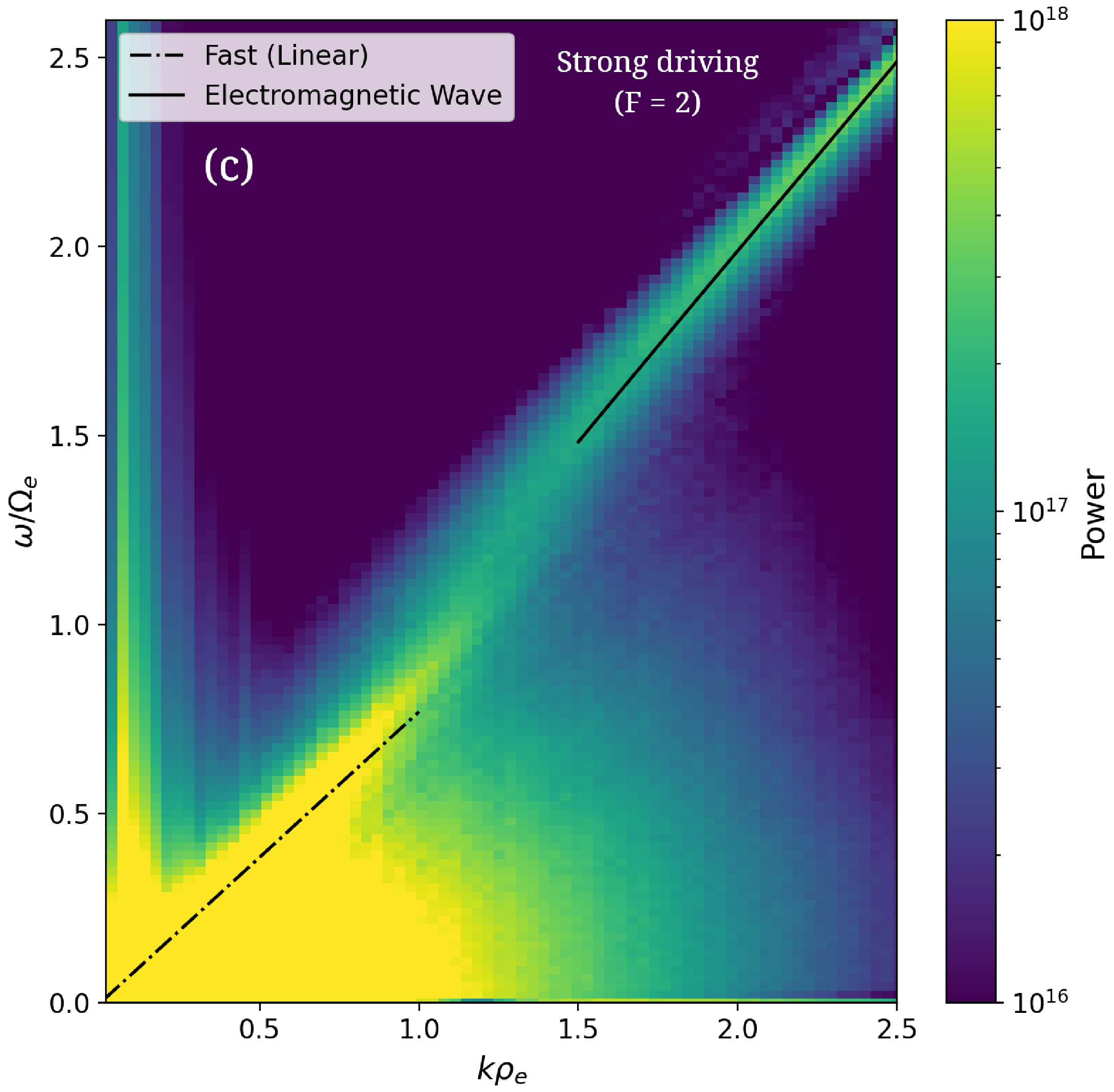}
  \caption{Spatiotemporal Fourier spectra in the $\omega$ vs. $k_\perp$ domain for three different quantities: (a) magnetic field spectrum $P_B$,  (b) compressive flow velocity spectrum $P_{u_c}$, and (c) strong driving ($F = 2$) magnetic field spectrum $P_{B}$.}
  \label{fig:temporalspectra}
\end{figure*}
Fig.~\ref{fig:hedgehog} shows 2D spatial Fourier power spectra of the magnetic field for weak and strong driving. In the weakly driven case ($F = 0.25$, panel (a)), the spectrum is organized into narrow, radially-extended beams in $k$-space, indicating that the FM turbulence in this regime is only weakly dispersive, the spectral power following a discrete set of linear dispersion relations. In contrast, the strongly driven case ($F = 2$, panel (b)) exhibits a substantially smoother distribution of spectral power, which indicates that nonlinear interactions and dispersive effects are no longer perturbative. This qualitative weak-strong transition resembles the scenario established for 3D acoustic turbulence \cite{Kochurin2024}. The presence of spectral beams in the weak-driving case here suggests an analogous weakly nonlinear FM cascade, in which energy transfer proceeds primarily through resonant wave–wave interactions.

Thus, to isolate subsonic FM turbulence (rather than shock turbulence), we next focus on a fiducial simulation in the weak driving regime, having $F = 0.25$ and $L/\rho_e = 256$. We use a spatiotemporal Fourier transform \cite{Gan2022,Arro2025, Fu2022} to compute the $\omega$ vs. $k_\perp$ power spectrum in 2D for the fiducial simulation over the time interval $t = 3 L/c$ to $4.1 L/c$, with a cadence of field dumps equal to $0.0045 L/c$ for a total number of 246 snapshots. Figure~\ref{fig:temporalspectra} shows several spatiotemporal power spectra: the magnetic field spectrum $P_B = \int_0^{2\pi} d\phi \, k_\perp B(\omega,\boldsymbol{k}_\perp)^2$, the compressive bulk flow velocity spectrum $P_{u_c} = \int_0^{2\pi} d\phi \, k_\perp |\boldsymbol{u}_c(\omega, k_\perp)|^2$, and $P_B$ for the strong driving case with F = 2, where the angle $\phi$ is the polar angle in the $x-y$ plane. We numerically approximate these integrals as sums. The solenoidal bulk flow power spectrum $P_{u_s} = \int_0^{2\pi} d\phi \,k_\perp |\boldsymbol{u}_s(\omega, k_\perp)|^2$ is subdominant, as shown in the Supplemental Material \cite{SuppMaterial}. Here, the bulk velocity field $\boldsymbol{u}=\boldsymbol{u}_c+\boldsymbol{u}_s$ is Helmholtz decomposed into a compressive part $\boldsymbol{u}_c$ and solenoidal part $\boldsymbol{u}_s$, satisfying $\nabla \times \boldsymbol{u}_c = 0$ and $\nabla \cdot \boldsymbol{u}_s = 0$ \cite{arro2025nature}. To reduce PIC noise in the high wavenumber region of these spectra, we subtracted the spatiotemporal spectrum of noise, obtained from a simulation run with the same physical parameters but no driving, so that $P = |P_{\rm orig} - P_{\rm noise}|$, where $P_{\rm orig}$ is the original spatiotemporal spectrum and $P_{\rm noise}$ is the spatiotemporal spectrum for the non-driven simulation. The spatiotemporal spectrum of noise $P_{\rm noise}$ for this case can be seen in Supplemental Material \cite{SuppMaterial}. PIC noise can act as a seed for the excitation of electromagnetic waves at high $k$, and we observe that the spectra of noise simulations show significant unphysical power along the electromagnetic wave branch. Thus, we remove these modes by subtracting the spectra. In panel (a) of Figure~\ref{fig:temporalspectra}, we can see that $P_B$ has good agreement with the analytical FM dispersion relation (Eq. \ref{eq:dispersion}, plotted as a dashed line), without nonlinear corrections in $k_\perp \rho_e$.  Interestingly, the spectral power at kinetic scales does not follow the higher order quadratic correction (in $k_\perp \rho_e$) to the analytical dispersion relation, which is surprising given that next order terms are expected to become significant at $k_\perp \rho_e \sim 1$; instead, the leading order FM dispersion relation is recovered even at $k_\perp \rho_e \gtrsim 1$. However, the spectral power in the FM is damped at high $k_\perp \rho_e \gg 1$. Panel (b) shows that $P_{u_c}$ also follows the FM dispersion relation (dashed line). We observe that there is a spike in power at $k_\perp \approx 2\pi/L$ in panels (a) and (b), which is associated with the energy injected by the external driving. Panel (c) shows that $P_B$ for the strongly driven case exhibits a clear transition from the fast mode dispersion relation to the electromagnetic mode at kinetic scales, which survives the noise subtraction. This spectrum also shows significant non-wave low-frequency power associated with nonlinear wave interactions in the strong driving regime. 

In FM turbulence,  
$B/B_0 = n/n_0$ \cite{Passot2003}. This follows from Faraday's Law in our 2D setup if we assume the ideal MHD electric field $\boldsymbol{E}=-\boldsymbol{u}\times\boldsymbol{B}$; then $\boldsymbol{B} = B \hat{\boldsymbol{z}}$ evolves via the continuity equation, mirroring density:
\begin{eqnarray}
    \frac{\partial B}{\partial t} = -c \hat{\boldsymbol{z}} \cdot \nabla \times \boldsymbol{E} = - c \nabla \cdot (\boldsymbol{u} B) \, .
\end{eqnarray}
Assuming $B = B_0 + \delta B$ and $|\delta B| \ll B_0$, to linear order the magnetic field fluctuations obey
\begin{eqnarray}
\delta B(k_\perp) = \frac{B_0}{\omega(k_\perp)} \boldsymbol{k_\perp} \cdot \boldsymbol{u} \sim \frac{B_0}{v_F} |\boldsymbol{u}| \, ,
\end{eqnarray}
where we used that $\omega(k_\perp) \sim v_F k_\perp$ and $|\nabla \times \boldsymbol{u}| \ll  |\nabla \cdot \boldsymbol{u}|$, as justified by the spatiotemporal spectra. This implies $|\boldsymbol{B}(k_\perp)|^2 \sim |\boldsymbol{u}(k_\perp)|^2$ at fluid scales. In addition, again using $\boldsymbol{E} = - \boldsymbol{u} \times \boldsymbol{B}$ and $|\delta B| \ll B_0$ gives
\begin{equation}
|\boldsymbol{E}|^2 = \left|\frac{\boldsymbol{u}}{c} \times \boldsymbol{B}\right|^2 \sim \frac{B_0^2}{c^2} |\boldsymbol{u}|^2 \, .
\end{equation}
These considerations imply that $\boldsymbol{B}, \: \boldsymbol{u}, \: \boldsymbol{E}, \: n$ must all have the same power-law spectral indices in the inertial range.

\begin{figure*}
  \centering
  \includegraphics[width=0.32\textwidth]{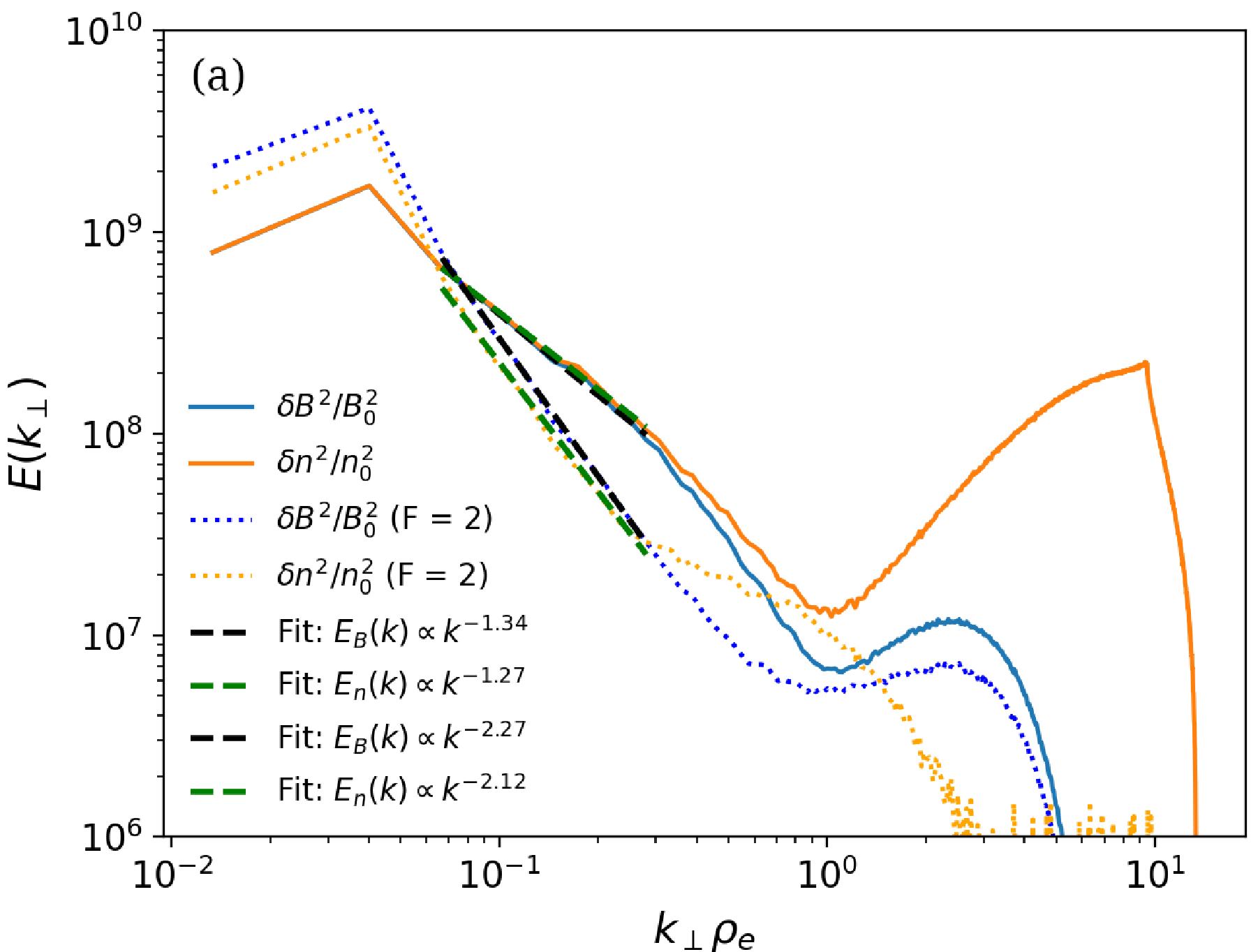}
  \hfill
  \includegraphics[width=0.32\textwidth]{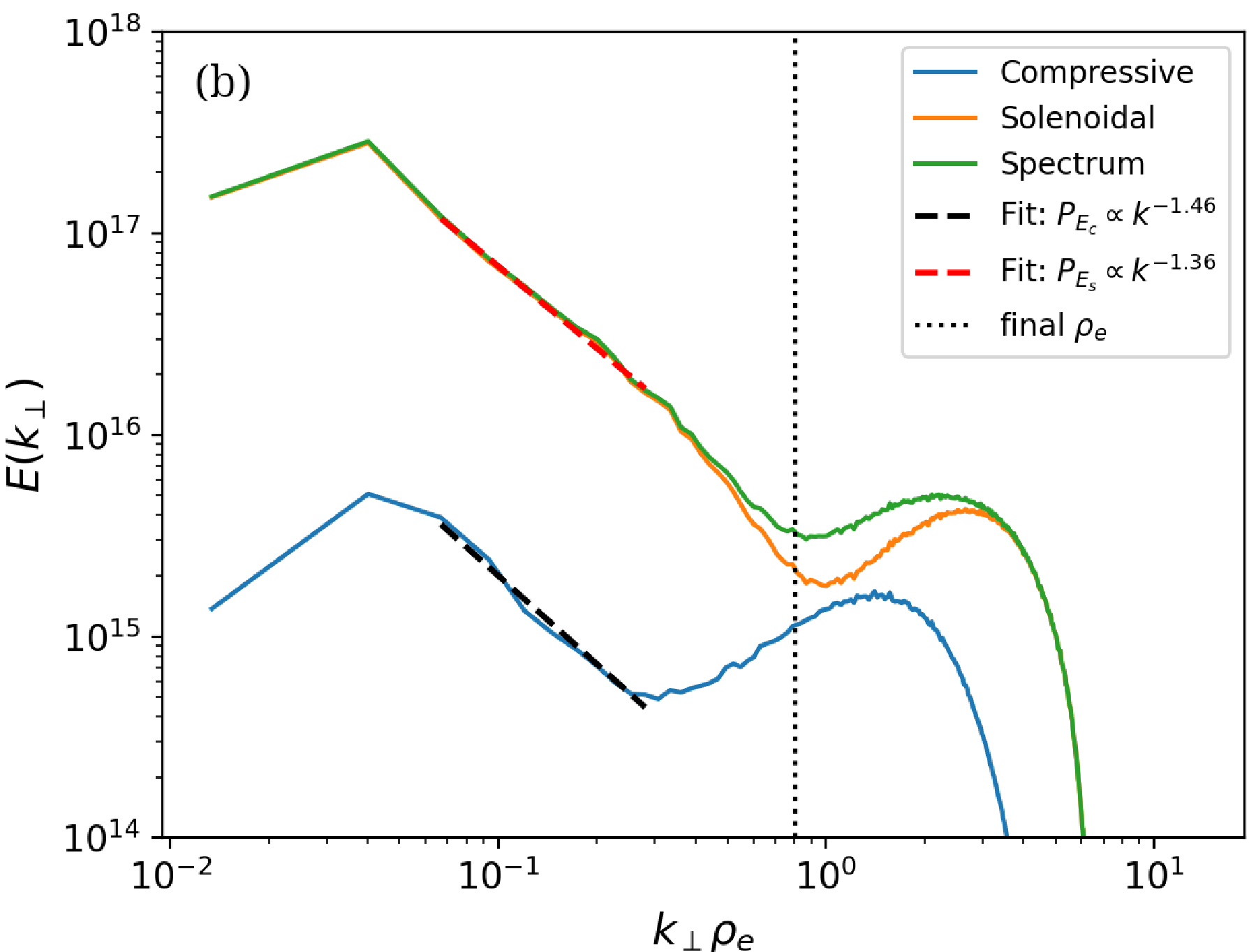}
  \hfill
  \includegraphics[width=0.32\textwidth]{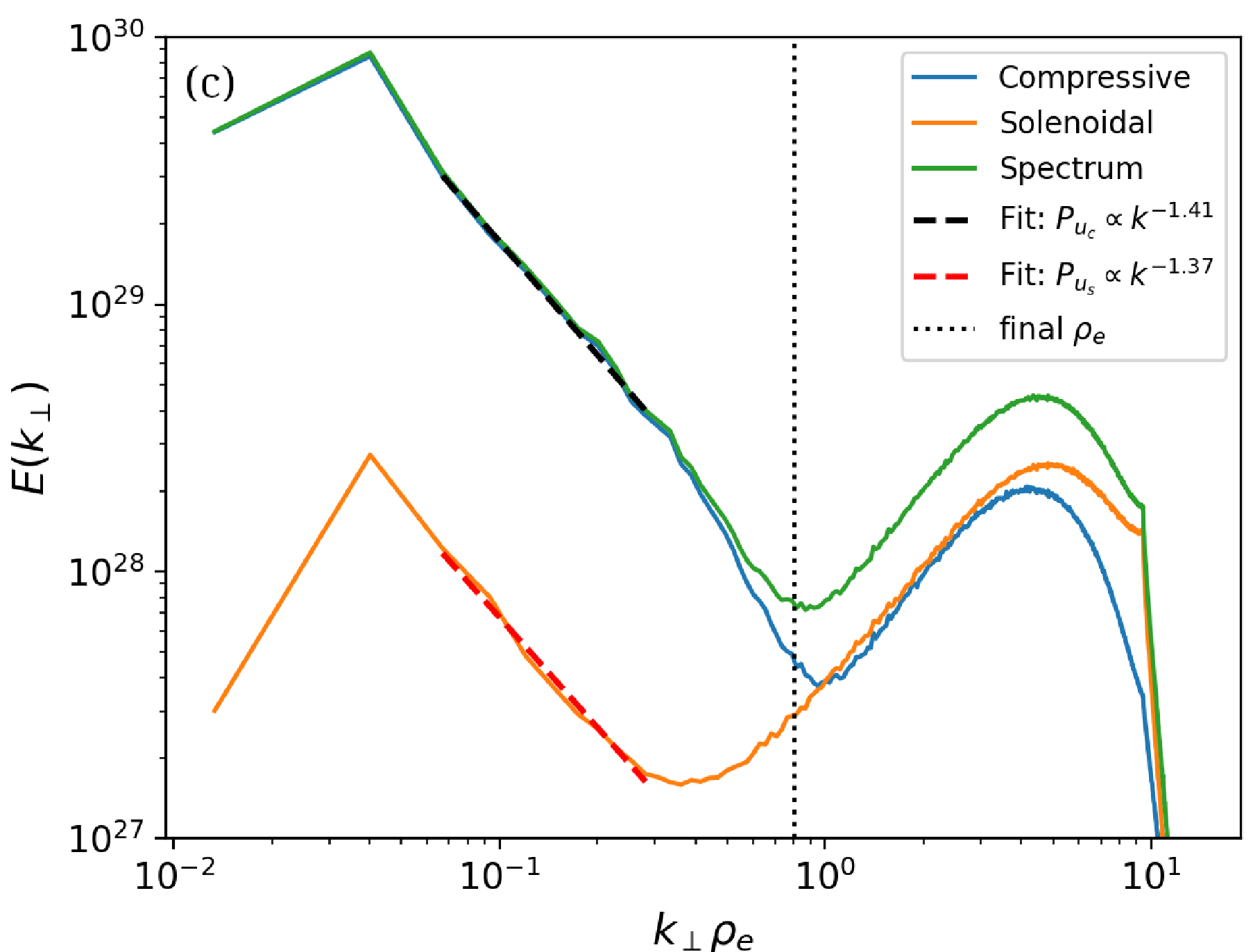}
  \caption{Spatial spectra for four different quantities: (a) Magnetic field $P_B(k_\perp)$ (blue) and density $P_n(k_\perp)$ (orange) spectra for both $F = 0.25$ (solid) and $F = 2$ (dotted). (b) Electric field $P_E(k_\perp)$ spectrum (green) along with its compressive $P_{E_{c}}(k_\perp)$ (blue) and solenoidal $P_{E_{s}}(k_\perp)$ (orange) components. (c) Flow velocity $P_u(k_\perp)$ spectrum along with its compressive $P_{u_{c}}(k_\perp)$ and solenoidal $P_{u_{s}}(k_\perp)$ components.}
  \label{fig:spatialspectra}
\end{figure*}

The (noise-subtracted) spatial Fourier power spectra for $B$, $\boldsymbol{u}$, $\boldsymbol{E}$, and $n$ are shown in Fig.~\ref{fig:spatialspectra}, where we plot $P_B(k_\perp)$, $P_n (k_\perp)$, $P_E (k_\perp)$, and $P_u(k_\perp)$. As predicted, these four spectra all have similar power-law indices in the inertial range, which we find to be near $-1.35$. For stronger driving ($F \sim 1$), we expect the Burgers turbulence result of a power law index of $-2$ as the turbulence becomes shock-dominated. This can be seen in panel (a), where the dotted lines show $P_B(k_\perp)$ and $P_n(k_\perp)$ for $F = 2$, which have an inertial-range spectral index of around $-2.2$. Panels (b) and (c) of Fig.~\ref{fig:spatialspectra} also include the Helmholtz decomposition of both $\boldsymbol{E}$ and $\boldsymbol{u}$, respectively. Most of the power is in the compressive component of $\boldsymbol{u}$, as expected. In contrast, we observe that most of the power is in the solenoidal component of $\boldsymbol{E}$, implying that the electric field is mostly electromagnetic rather than electrostatic, as expected for FMs.

Obtaining a theoretical prediction for the magnetic energy spectrum for in 2D FM turbulence is challenging. The linear FM dispersion relation takes a similar form to acoustic sound waves. For 2D dispersionless acoustic turbulence, weak wave turbulence theory fails due to the fast growth of secular terms, leading to a singular integral in the wave kinetic equation \cite{Galtier2023}. In 3D, the growth of secular terms is much slower, allowing the singularity to be integrable \cite[see][]{Galtier2023}. 

The 2D problem is more tractable if the acoustic wave is weakly dispersive with the form $\omega(k_\perp) = c_s k_\perp [1 + (ak_\perp)^2]$, where $a \ll 1/k_\perp$ is a dispersion length scale. In this case, a spectrum of $P_u(k_\perp) \propto k_\perp^{-1}$ was derived \cite{Griffin2022}. Unfortunately, the dispersive FM from Eq.~\ref{eq:dispersion} has a negative dispersion, for which three-wave interactions do not admit resonance. 

{\it Discussion/Conclusions--- } In this Letter, we present the results of novel 2D PIC simulations of compressively driven pair plasma, demonstrating FM wave turbulence. We find that weak compressive driving is effective for establishing a FM cascade. Using spatiotemporal Fourier analysis, we identify a clear concentration of power along the analytically predicted FM dispersion relation across a broad range of scales, extending from fluid to kinetic regimes. The persistence of linear FM power to kinetic scales is particularly curious: rather than transitioning to a FM with kinetic corrections or a high-frequency Langmuir wave, a substantial fraction of power in magnetic/velocity spectra remains along the linear MHD FM branch. We hypothesize that this might be due to effective collisionality near the kinetic scales, making the FM more fluid-like than naively expected.

A key result is that, in the weakly driven regime, the magnetic, electric, velocity, and density fields exhibit nearly identical inertial-range spatial spectra with a power-law index close to $k_\perp^{-4/3}$. This motivates further theoretical studies regarding the scaling of relativistic FM turbulence.

An important outcome of this work is the demonstration that FM turbulence can persist as a wave-dominated cascade in a fully kinetic plasma, without inevitably steepening into shocks. Although FMs are often associated with shock formation in MHD turbulence studies (e.g., \citep{ChoLazarian2002, ChoLazarian2003}), and in hybrid-kinetic simulations at high Mach number \citep{gootkin_etal_2025}, we find that in the weakly driven regime nonlinear steepening is suppressed and the dynamics are governed by sustained wave–wave interactions. This is evidenced by the small supersonic fraction and the concentration of spectral power along the linear FM dispersion relation across fluid and kinetic scales. Our results therefore establish that wave turbulence remains a valid and self-consistent description of FM dynamics in collisionless, relativistic plasmas.

In our 2D setup, standard energy dissipation channels such as magnetic reconnection, parallel Landau damping, and viscous stress from pressure-anisotropy microinstabilities are absent. Remaining channels of damping may include cyclotron resonance, stochastic heating \cite{Chandran2010}, and shocks. Shocks can be ruled out for our simulations in the weak driving regime. Similarly, stochastic heating is inhibited in the weak turbulence regime where $\delta B/B_0 \ll 1$. This leaves cyclotron damping of FMs as the most viable dissipation channel, which can cause irreversible heating even in the absence of explicit collisions. As the driving amplitude increases, the supersonic fraction grows and the system transitions toward a shock-dominated regime, where the cascade steepens toward a Burgers-like $k_\perp^{-2}$ spectrum. This marks a qualitative change in the dissipation pathway: from quasilinear, resonant heating in the weakly nonlinear regime to strongly nonlinear dissipation in shocks.

A detailed analysis of dissipative processes and entropy production is deferred to future work, which will build on these simulation results. This work may also have astrophysical applications, a notable example being cosmic-ray scattering in the interstellar medium where FMs provide the dominant pitch-angle scattering contribution \citep{Yan2004}. More broadly, we speculate that similar weak turbulence phenomena may occur in the long-wavelength limit of magnetized electron-hole gases within 2D lattice materials such as graphene, which would provide a laboratory analogue of our system \citep{kotov_etal_2012}.

The authors thank Giorgio Krstulovic for insightful comments on acoustic turbulence with negative dispersion, and Karol Fulat for fruitful discussions. The authors acknowledge support from the National Science Foundation under NSF grant PHY-2409316, and the Department of Energy under the grant DE-SC0026099. This work used Stampede3 at the Texas Advanced Computer Center (TACC) through allocation PHY160032 from the Advanced Cyberinfrastructure Coordination Ecosystem: Services \& Support (ACCESS) program, which is supported by U.S. National Science Foundation grants \#2138259, \#2138286, \#2138307, \#2137603, and \#2138296. 

\bibliography{main}

\end{document}


\title{Supplemental Material}

\maketitle

\section{Noise Simulation Spatiotemporal Spectrum}

\begin{figure}
\centering
\includegraphics[width=0.49\textwidth]{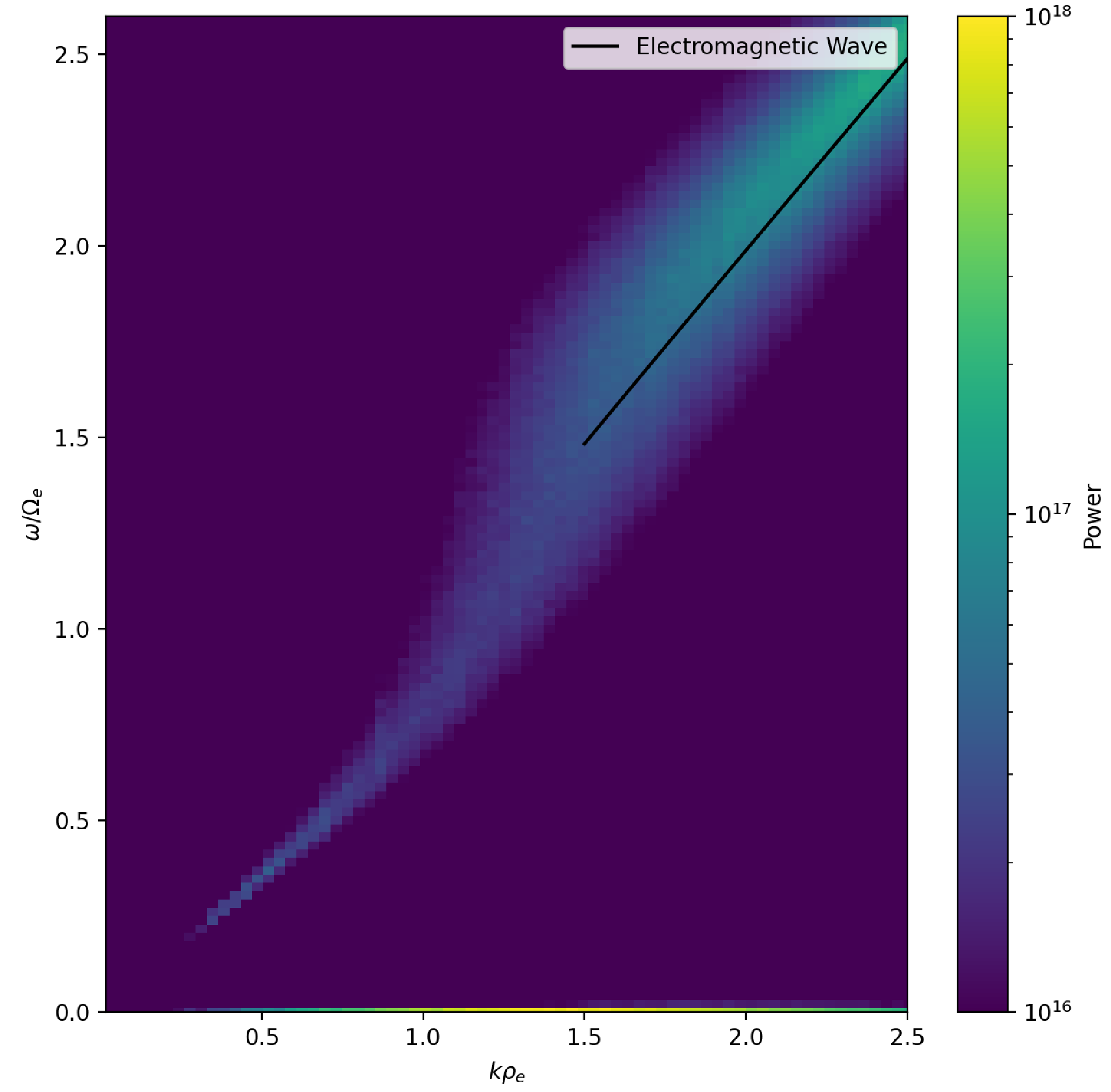}
\hfill
\includegraphics[width=0.49\textwidth]{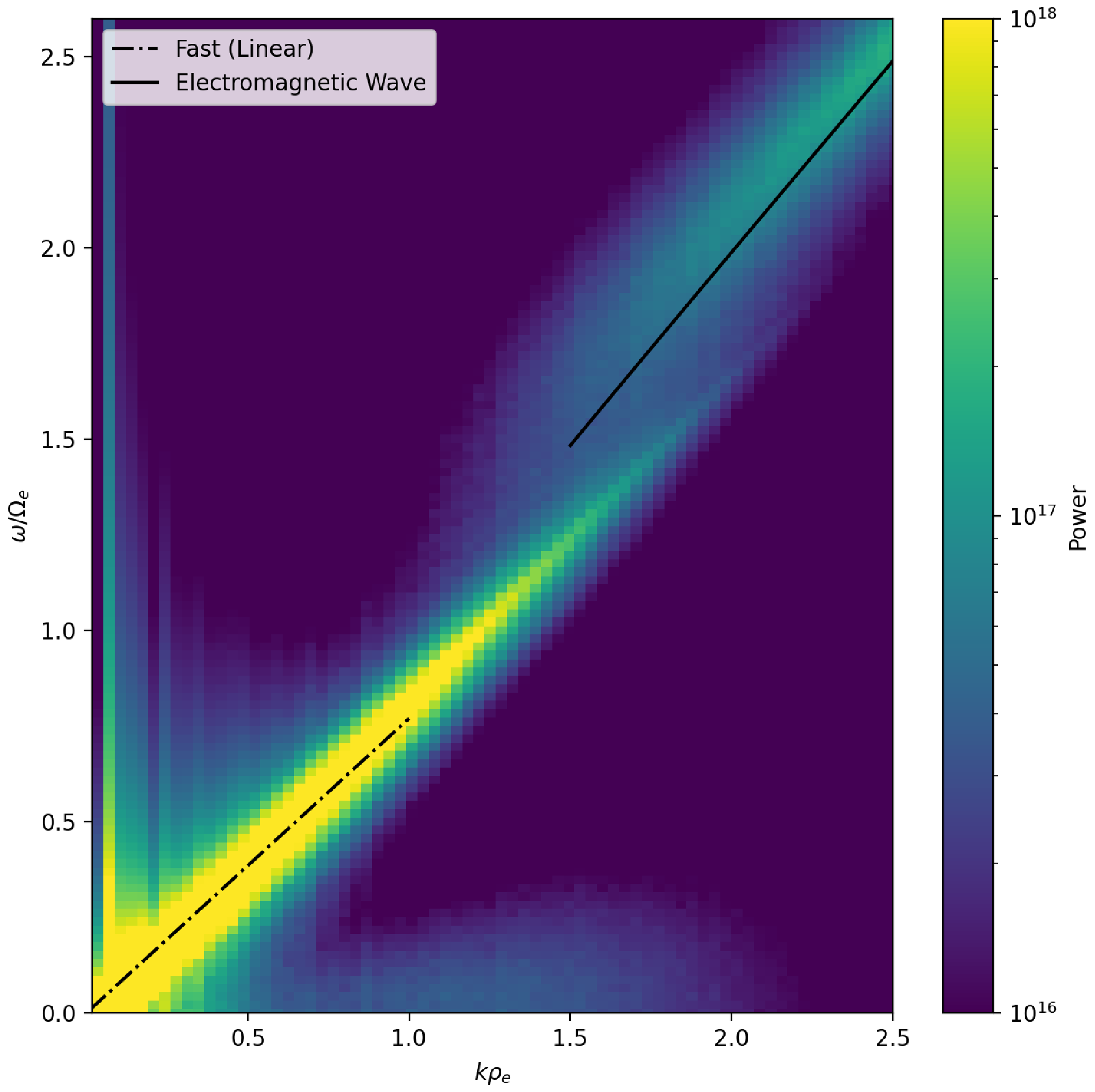}
\caption{Spatiotemporal Fourier spectra for the magnetic field; the left panel shows $P_\text{noise}(k_\perp, \omega)$ for the noise simulation and the right panel shows $P_\text{original}(k_\perp, \omega)$ for the original fiducial turbulence simulation without noise subtraction.}
\label{fig:noise}
\end{figure}

In Fig.~\ref{fig:noise}, we show the spatiotemporal Fourier spectra of the noise simulation (left) and the turbulence simulation  without noise subtraction (right), referenced in the main text as the fiducial case. Note that in the noise simulation for $k_\perp \rho_e \gtrsim 1$, the spectral power becomes strongly concentrated along a well-defined ridge that closely follows the electromagnetic wave dispersion relation (indicated by the black line). This alignment demonstrates that, even in the absence of external driving ($F = 0$), the simulation supports spontaneously excited propagating electromagnetic modes. At lower $k_\perp \rho_e \lesssim 1$, the spectral power in the noise simulation is negligible, as expected since PIC noise fluctuations occur predominantly at kinetic scales.

\section{Solenoidal Bulk Flow Velocity Power Spectrum}

\begin{figure}
\includegraphics[width=0.49\textwidth]{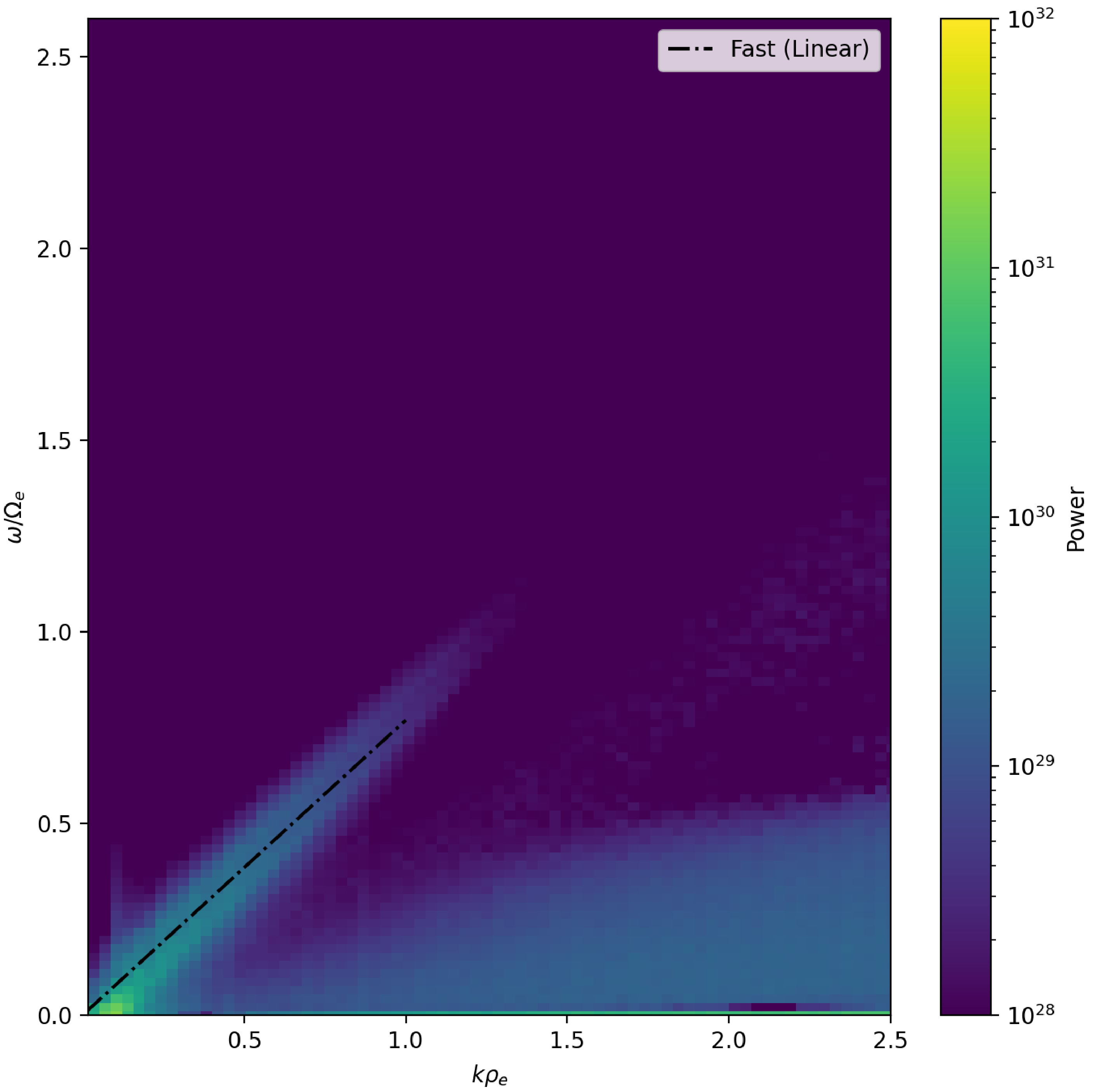}
\caption{Spatiotemporal Fourier spectrum $P_{u_{s}}(k_\perp, \omega)$ for the solenoidal flow velocity in the fiducial turbulence simulation.}
\label{fig:solenoidal}
\end{figure}

Fig.~\ref{fig:solenoidal} shows the solenoidal power spectrum of the bulk flow velocity $P_{u_{s}} = k_\perp |\boldsymbol{u}_s(k_\perp, \omega)|^2$ for the spatiotemporal simulation with $L/\rho_e = 256$ and $F = 1$. In contrast to the compressive power (shown in the main paper), the solenoidal power is much smaller and is concentrated at lower frequencies with a broad distribution in $k$. This is characteristic of non-wave motions, as it does not follow any linear dispersion relation \citep{papini2021spacetime,Gan2022,Arro2025,arro2025nature}. However, there is some residual structure along the FM dispersion relation. 

\section{Derivation of Dispersion Relation}

Following Eq.~(10-48) of Stix \cite{Stix1992}, the general susceptibility tensor for an arbitrary gyrotropic distribution function $f_{0,s}(p_\perp,p_\parallel)$  for species $s$ can be written as:
\begin{align}
\boldsymbol{\chi}_s
= \frac{\omega_{p0,s}^2}{\omega \Omega_{0,s}}
\int_0^\infty 2\pi p_\perp \, dp_\perp
\int_{-\infty}^{\infty} dp_\parallel
& \left[ \hat{\mathbf e}_\parallel \hat{\mathbf e}_\parallel
\frac{\Omega_s}{\omega}
\left(
\frac{1}{p_\parallel}\frac{\partial f_{0,s}}{\partial p_\parallel}
-
\frac{1}{p_\perp}\frac{\partial f_{0,s}}{\partial p_\perp}
\right)
p_\parallel^2 + \sum_{n=-\infty}^{\infty}
\frac{\Omega_s p_\perp U_s}
{\omega - k_\parallel v_{\parallel,s} - n\Omega_s}
\, \mathbf{T}_n\left(\frac{k_\perp v_{\perp,s}}{\Omega_s}\right) \right]\, ,
\label{eq:chi_general}
\end{align}
where $\omega_{p0,s} = \sqrt{4\pi n_s q_s^2/m_s}$ and $\Omega_{0, s} = q_sB/m_sc$ refer to the non-relativistic plasma frequency and gyrofrequency, where $n_s, q_s, m_s$ are the density, charge, and mass of particles of species $s$. Components of vectors parallel and perpendicular to the magnetic field are denoted by the subscripts $\parallel$ and $\perp$, respectively. The velocity vector is then defined by $\boldsymbol{v}_s = \boldsymbol{p}/m_s\gamma_s$, where $\gamma_s = \sqrt{1 + \boldsymbol{p}^2/m_s^2 c^2}$ is the Lorentz factor. In addition, in the susceptibility,
\begin{align}
U_s
&=
\frac{\partial f_{0,s}}{\partial p_\perp}
+
\frac{k_\parallel}{\omega}
\left(
v_{\perp,s} \frac{\partial f_{0,s}}{\partial p_\parallel}
-
v_{\parallel,s} \frac{\partial f_{0,s}}{\partial p_\perp}
\right),
\label{eq:U_def}
\\
\Omega_s
&=
\frac{\Omega_{0,s}}{\sqrt{1 + \frac{\boldsymbol{p}^2}{m_s^2 c^2}}}
=
\frac{\Omega_{0,s}}{\gamma_s}.
\label{eq:Omega_def}
\end{align}
The tensor $\mathbf{T}_n(z)$ is given by
\begin{align}
\mathbf{T}_n =
\begin{pmatrix}
\displaystyle
\frac{n^2 J_n^2(z)}{z^2}
&
\displaystyle
\frac{i n J_n(z) J_n'(z)}{z}
&
\displaystyle
\frac{n J_n^2(z) p_\parallel}{z p_\perp}
\\[1.2ex]
\displaystyle
-\frac{i n J_n(z) J_n'(z)}{z}
&
\displaystyle
\left[J_n'(z)\right]^2
&
\displaystyle
-\frac{i J_n(z) J_n'(z) p_\parallel}{p_\perp}
\\[1.2ex]
\displaystyle
\frac{n J_n^2(z) p_\parallel}{z p_\perp}
&
\displaystyle
\frac{i J_n(z) J_n'(z) p_\parallel}{p_\perp}
&
\displaystyle
\frac{J_n^2(z) p_\parallel^2}{p_\perp^2}
\end{pmatrix},
\label{eq:Tn_def}
\end{align}
where $z = k_\perp v_{\perp,s}/\Omega_s$, and $J_n(z)$ are modified Bessel functions of order $n$.

Assuming an isotropic equilibrium distribution $f_0(p)$ and purely perpendicular propagation ($k_\parallel=0$), we may choose $\boldsymbol{k}=k_\perp \hat{\boldsymbol{x}}$ without loss of generality (by axisymmetry about $\boldsymbol{B}_0$). In this geometry the electromagnetic fast-mode branch corresponds to the extraordinary (X) polarization with $E_y\neq 0$, so the relevant response is the $yy$ component of the susceptibility, $\chi_{yy}$, which simplifies to:
\begin{align}
\chi_{yy_s} = \frac{2\pi \omega_{p0,s}^2}{\omega \Omega_{0,s}}  \int_0^\infty p_{\perp}^2 dp_\perp \int_{-\infty}^{\infty} dp_\parallel \sum_{n=-\infty}^{\infty} \frac{\Omega_{s}}{\omega-n\Omega_{s}} \frac{\partial f_{0,s}}{\partial p_\perp} J'^2_n\left(\frac{k_\perp p_\perp}{m_s \Omega_{0,s}}\right)
\end{align}

Summing over species $s$, assuming a pair plasma (with identical background distributions for both species $f_0$):
\begin{align}
\chi_{yy} = \frac{2\pi \omega_{p0,e}^2}{\omega}  \int_0^\infty p_{\perp}^2 dp_\perp \int_{-\infty}^{\infty} dp_\parallel \left[ \frac{2}{\gamma \omega} \frac{\partial f_0}{\partial p_\perp}J'^2_0\left(\frac{k_\perp p_\perp}{m \Omega_{0,e}}\right) + \sum_{n \neq 0} \frac{2\gamma \omega}{\gamma^2 \omega^2-n^2\Omega_{0,e}^2} \frac{\partial f_0}{\partial p_\perp} J'^2_n\left(\frac{k_\perp p_\perp}{m \Omega_{0,e}}\right)\right]
\end{align}

\subsection{Low-frequency and long-wavelength expansion 
($\omega/\Omega_e \ll 1,\; k\rho_e \ll 1$)}

Assuming $\omega/\Omega_{0,e} \ll 1$ and $k_\perp \rho_e \ll 1$, implying that orders $|n| = 0, 1, 2$ are dominant (to second order in these approximations):
\begin{align}
\chi_{yy} = \frac{2\pi \omega_{p0,e}^2}{\omega}  \int_0^\infty p_{\perp}^2 dp_\perp \int_{-\infty}^{\infty} dp_\parallel \frac{\partial f_0}{\partial p_\perp} \left[ \frac{2}{\gamma  \omega} J'^2_0\left(\frac{k_\perp p_\perp}{m \Omega_{0,e}}\right)  - \left(\frac{4 \gamma \omega}{\Omega_{0,e}^2} + \frac{4 \gamma^3 \omega^3}{\Omega_{0,e}^4}\right)J'^2_1\left(\frac{k_\perp p_\perp}{m \Omega_{0,e}}\right) - \frac{ \gamma \omega}{\Omega_e^2} J'^2_2\left(\frac{k_\perp p_\perp}{m \Omega_{0,e}}\right)\right] \end{align}
Nondimensionalizing with $\zeta_\perp = k_\perp p_\perp/m\Omega_{0,e}$, and $\zeta_\parallel = k_\perp p_\parallel/m\Omega_{0,e}$, we can expand the Bessel functions assuming $k_\perp\rho_e \ll 1$, with an ultrarelativistic Lorentz factor $\gamma \approx |\boldsymbol{p}|/mc \gg 1$:
\begin{align}
\chi_{yy} 
&= 2\pi \omega_{p0,e}^2 m^3 \Omega_{0,e}^3 k_\perp^{-3}
\int_0^\infty \zeta_{\perp}^2 \, d\zeta_\perp 
\int_{-\infty}^{\infty} d\zeta_\parallel 
\frac{\partial f_0}{\partial \zeta_\perp}  \\
&\quad \times \Bigg[
 \frac{k_\perp c}{ 2\Omega_{0,e} \omega^2}
\frac{\zeta_\perp^2}{\sqrt{\zeta_\parallel^2 + \zeta_\perp^2}}
- \frac{k_\perp c}{ 8\Omega_{0,e} \omega^2}
\frac{\zeta_\perp^4}{\sqrt{\zeta_\parallel^2 + \zeta_\perp^2}}- \frac{4}{k_\perp c\Omega_{0,e}}
\sqrt{\zeta_\parallel^2 + \zeta_\perp^2}
\left(
\frac{1}{4}
- \frac{3 }{16}\zeta_\perp^2
\right)
\\
&\quad
- \frac{\omega^2}{k_\perp^3 c^3 \Omega_{0,e}}
(\zeta_\parallel^2 + \zeta_\perp^2)^{3/2}
 - \frac{9 }{144  k_\perp c \Omega_{0,e}}
\sqrt{\zeta_\parallel^2 + \zeta_\perp^2}  \zeta_\perp^2
\Bigg]
\end{align}
At this step, the form of the distribution function is needed to proceed in evaluating the integrals. We choose the ultra-relativistic Maxwell--J\"uttner distribution $f(\zeta_\perp, \zeta_\parallel) = (1/8\pi m^3 \theta^3 c^3) e^{-(\Omega_{0,e}/k_\perp c\theta)\sqrt{\zeta_\parallel^2 + \zeta_\perp^2}}$, evaluating the integrals gives:
\begin{align}
\chi_{yy} = - \frac{16\omega_{p0,e}^2  \theta c^2 }{5 \omega^2 \Omega_{0,e}^2 }k_\perp^2 + \frac{144\omega_{p0,e}^2  \theta^3 c^4}{7  \omega^2 \Omega_{0,e}^4 }k_\perp^4   + \frac{8\omega_{p0,e}^2 \theta}{\Omega_{0,e}^2} -\frac{132\omega_{p0,e}^2 \theta^3 c^2}{ \Omega_{0,e}^4} k_\perp^2  +\frac{240 \omega_{p0,e}^2 \theta^3}{\Omega_{0,e}^4}\omega^2 
\end{align}

Solving for the dispersion relation from the susceptibility and rewriting in terms of relativistic variables, we obtain the same form as in the main paper:
\begin{align}
\omega(k) = v_Fk_\perp \left(1 - A \frac{c^2}{\Omega_e^2} k_\perp^2\right), \text{ where }A \equiv \frac{ 7/10 + (7/5) v_A^2/c^2 + (5/2) v_A^4/c^4}{3(1+v_A^2/c^2)^2 (v_A^2/c^2 + 2/5)} 
\end{align}

\subsection{High-frequency and short-wavelength expansion 
($\omega/\Omega_e \gg 1,\; k\rho_e \gg 1$)}

Starting from (6), and expanding in the opposite limit $\omega/\Omega_{0,e} \gg 1$ and $k_\perp\rho_e \gg 1$, we obtain:
\begin{align}
\chi_{yy} = -\frac{\omega_{p0,e}^2}{16\theta \omega^2} + \frac{\omega_{p0,e}^2 \Omega_{0,e}^2}{8\omega^2\theta^3 c^2 k_\perp^2}
\end{align}

This gives the electromagnetic wave dispersion relation (now written with relativistic $\omega_{pe}$):
\begin{align}
\omega(k) = ck_\perp \left(1 + \frac{3\omega_{pe}^2}{32 c^2 k_\perp^2}\right) 
\end{align}

\section{Density Animations}

\textbf{Movie 1:} Density animation of $F = 0.25$ weak driving $(L/\rho_e = 512)$

\textbf{Movie 2:} Density animation of $F = 4$ strong driving $(L/\rho_e = 512)$

\bibliography{supplement}
